\documentclass[%
 reprint,
superscriptaddress,
 amsmath,amssymb,
aps,
prb,
]{revtex4-2}

\usepackage{graphicx}
\usepackage{dcolumn}
\usepackage{bm}
\usepackage{xcolor}

\newcommand{\ket}[1]{\left|#1\right>} 	

\newcommand{\av}[1]{\big<#1\big>}
\newcommand{\avt}[1]{\big<T_\text{c}\big(#1\big)\big>}

\definecolor{resonator-red}{HTML}{EA1F22}
\definecolor{resonator-green}{HTML}{03C028}
\definecolor{resonator-blue}{HTML}{0989ec}

\usepackage{lineno}
\usepackage{soul} 	
\usepackage[normalem]{ulem}

\begin{document}


\title{Quantum-correlated photons generated by nonlocal electron transport}

%
\author{Felicitas Hellbach}
\affiliation{Fachbereich Physik, Universit{\"a}t Konstanz, 78457 Konstanz, Germany}
\author{Fabian Pauly}
\affiliation{Institute of Physics, University of Augsburg, 86135 Augsburg, Germany}
\author{Wolfgang Belzig}
\affiliation{Fachbereich Physik, Universit{\"a}t Konstanz, 78457 Konstanz, Germany}
\author{Gianluca Rastelli}
\affiliation{Fachbereich Physik, Universit{\"a}t Konstanz, 78457 Konstanz, Germany}
\affiliation{INO-CNR BEC Center and Dipartimento di Fisica, Universit{\`a} di Trento, 38123 Povo, Italy}

\date{\today}

\begin{abstract}
Since the realization of high-quality  microwave cavities coupled to quantum dots, one can envisage the possibility to investigate the coherent interaction of light and matter in semiconductor quantum devices. 
Here we study a parallel double quantum dot device operating as single-electron splitter interferometer, with each dot coupled to a local photon cavity. 
We explore, how quantum correlation and entanglement between the two separated cavities are generated by the coherent transport of a single electron passing simultaneously through the two different dots. We calculate the covariance of the cavity occupations by use of a diagrammatic perturbative expansion based on Keldysh Green's functions to the fourth order
in the dot-cavity interaction strength, taking into account vertex diagrams. 
Furthermore, we demonstrate the creation of entanglement by showing that the classical Cauchy-Schwarz inequality is violated if the energy levels of the two dots are almost degenerate. 
For large level detuning or a single dot coupled to two cavities, we show that the inequality is not violated.
\end{abstract}

\maketitle


%
%
%
%

\section{\label{sec:level1}Introduction}

Nonlocality is a fundamental property of quantum mechanics that manifests itself in two main ways: as delocalization of a quantum particle in space according to its associated wave function (superposition), as correlations between spatially separated parts of a quantum system (entanglement). It is at the heart of quantum communication and computing in various physical implementations.

An intriguing example of quantum delocalization is interference in the motion of a single electron.
Quantum delocalized transport has been proven in nanodevices 
formed by two possible paths connecting an initial and final point, 
namely two electrical contacts playing the role of source and drain.
Examples are parallel double dots \cite{Holleitner:2001,Sigrist:2007,Hatano:2011bn}, operating as single-electron splitter interferometer, or the electronic Mach-Zehnder interferometer \cite{PhysRevB.82.155303}, operating with the edge states of two-dimensional (2D) quantum Hall systems \cite{Ji:2003ck}. 
Similarly to a photon in a Mach-Zehnder interferometer, an electron wave-function can split in two branches and then by recombining 
give rise to interference in the transmitted flux.
In  general, semiconducting single-electron devices form a unique playground to address 
nonlocal electron transport and quantum interference
\cite{Schuster:1997hh,Bocquillon:2013dp,Holleitner:2001,Sigrist:2007,Hatano:2011bn}.

Besides electron transport, quantum mechanics can be explored with high precision in the field of optics and photonics. In particular, microwave quantum photonics has made remarkable progress in the last decade.
In the circuit quantum electrodynamics (QED) architecture \cite{Blais:2021be}, a large variety of quantum states in an electromagnetic microwave resonator has been prepared and measured  \cite{Hofheinz:2008dq,Hofheinz:2009ba}. 
Moreover, using superconducting qubits or Josephson circuitry (Josephson parametric amplifier or wave-mixer), quantum entangled states of microwave photons have been realized in two spatially separated resonator cavities \cite{Wang:2011}, in two resonator modes of different frequency \cite{ZakkaBajjani:2011in,Nguyen:2012} as well as in propagating photons \cite{Eichler:2011,Flurin:2012,Lang:2013es}. 
%
More recently, an entangled pair of two-mode cat states has been realized in two microwave 
cavities \cite{Wang:2016bt} and a dc-biased Josephson junction was used to create two continuous entangled microwave beams \cite{Peugot:2021}.

Beyond superconducting circuits based on Josephson junctions, quantum dots realized in semiconducting nanostructures implement reliable and well-controlled qubits \cite{Petta:2005kn,Koppens:2006kz} with transition frequencies in the microwave domain and with the advantage of electric field control \cite{Nowack:2007du}. 
Quantum dots can now be readily coupled to microwave photon cavities, establishing the field of semiconductor hybrid QED \cite{Burkard:2020cb}, which provides a novel family of coherent quantum devices that combine electronic with photonic degrees of freedom on-chip \cite{Petersson:2012cv,Rossler:2015dk,Viennot:2015ir,Bruhat:2016br,Hagenmuller:2018bl,Trif:2019-1,Maisi:2021,Cottet:2020,Dymtruk:2016,Jincheng:2019,Bhandari:2021}.
The so-called strong coupling regime has been reached \cite{Mi:2016ex,Stockklauser:2017bq,Mi:2018ip} as well as the full microwave control and readout of the quantum dot qubits \cite{Scarlino-PRL-2019}.

Coupling quantum dots with quantum optical resonators adds a new dimension to the cavity and circuit QED, beyond the conventional paradigm of an atom coupled to a harmonic oscillator. This research line opens the path to exploring the correlations between charge transport and nonequilibrium, possibly quantum, regimes of localized electromagnetic radiation.
The corresponding hybrid devices are promising for implementing quantum transducers, in which single electrons control photonic quantum states in microwave cavities.

%
%
%
%

\section{\label{sec:level2} The system}
In this context, we analyze a parallel double quantum dot system, as shown in Fig.~\ref{fig:1}(a), each dot capacitively coupled to one of two separated microwave cavities of resonance frequencies $\omega_{a}$ and $\omega_{b}$, respectively.
The two dots are connected to a common left and right lead with the same hopping parameter $t$. 
We denote the coupling strength of the dot-cavity interaction by $\lambda$.

We study a spinless model, consisting of a single electron state in each dot, whose energy level is given by $\varepsilon_\text{a,b}=\bar{\varepsilon} \pm \Delta\varepsilon$ for the upper and lower dot, respectively. 
There is no direct tunneling between the dots. A priori, we cannot exclude the possibility that both dots are occupied simultaneously. {However, the tunnelling of electrons into the double-dot system is an uncorrelated event in our model due to the lack of electron-electron interactions. The process itself can therefore not generate quantum correlations between the two microwave cavities that go beyond the elementary single-electron tunneling process that we discuss in the following.

%
%
%
\begin{figure}[t]
\includegraphics[scale=0.2]{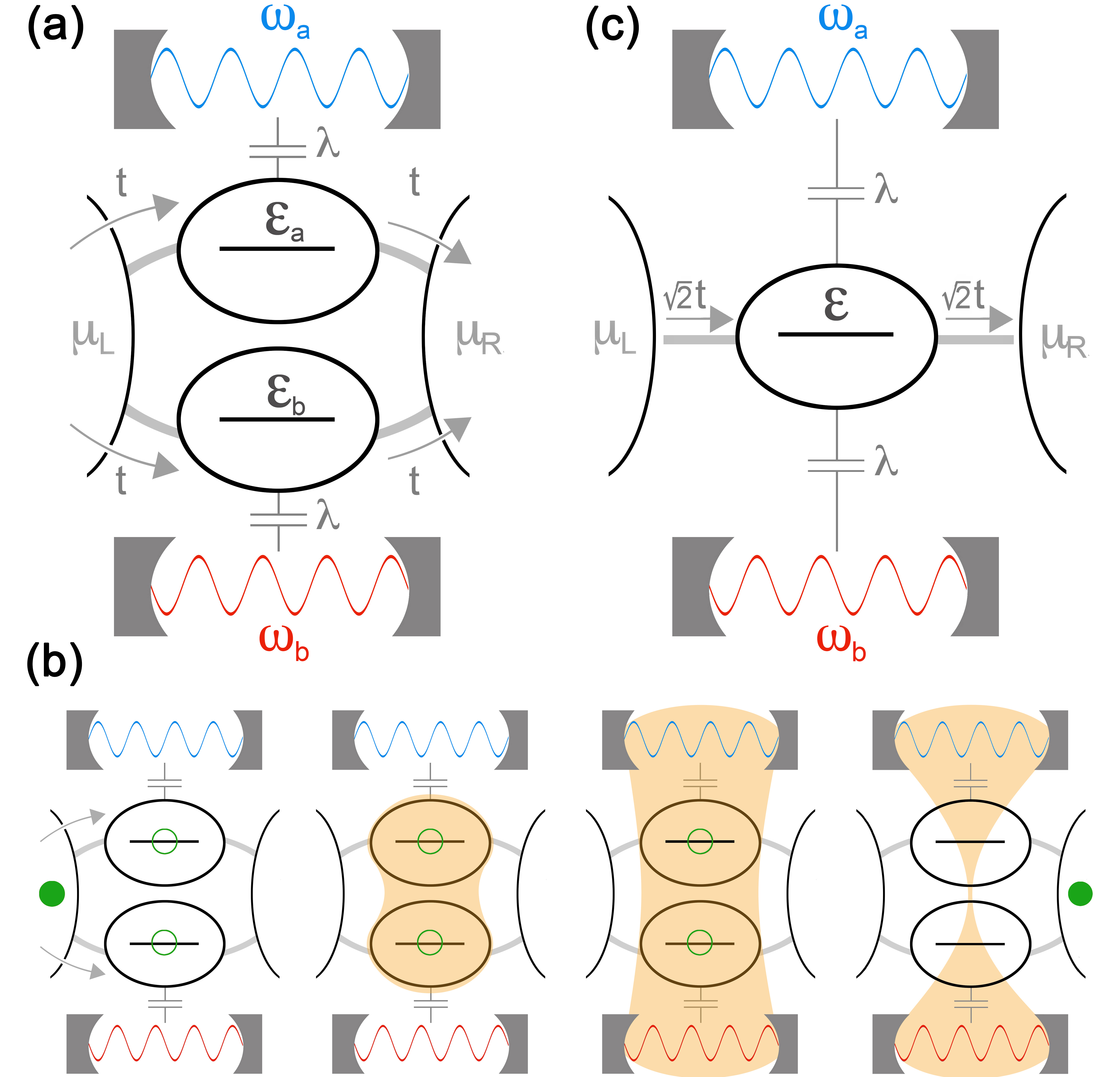}
\caption{\label{fig:1}
Sketch of studied systems and idea. (a) Main model: a parallel double quantum dot with a single electronic energy level in each quantum dot. Each dot is coupled to one of two separated microwave cavities and a common left and right lead. 
(b) Basic idea: An electron travels through both branches of the parallel double quantum dot simultaneously. Its state is a coherent superposition of the states in the two dots. 
The delocalization of the single electron yields quantum correlations between the two cavities and remains even when the electron leaves the system. (c)
Single quantum dot with a single electronic energy level coupled to two microwave cavities.}
\end{figure}

We explore how correlation and entanglement between the two cavities emerge by the coherent transport of a single electron passing simultaneously through the two different dots. 
Let us first assume that the energy levels of the dots are close to each other, in the sense that the energy difference is small compared to their broadening $|\Delta \varepsilon|=|\varepsilon_{\text{a}}-\varepsilon_{\text{b}}|\ll \Gamma$, i.e.\ the energy distributions overlap. 
In this regime the two paths are indistinguishable, and the electron flows through both branches simultaneously,
causing quantum interference in the double quantum dot system.
This means that the linear conductance associated with the levels of the double dot system is different from the sum of the linear conductances of the two separate dots, i.e.\ different from the single level regime.
If the difference between the two energy levels is increased, i.e.\ $|\Delta\varepsilon|\gg \Gamma$, the interference is destroyed, 
and the electron transfer occurs via the incoherent sum of the two possible paths, namely the electron proceeds independently through the upper or through the lower branch, but not simultaneously through both. This is the mechanism that allows or prevents entanglement. 

As illustrated in Fig.~\ref{fig:1}(b), the idealized procedure is as follows: When the electron travels inside the system, it splits. 
Therefore the electronic state is a coherent superposition of the electron occupying the upper dot or the lower dot, with the corresponding occupations $n^{(\text{el})}_\text{a}$ and $n^{(\text{el})}_\text{b}$.
The state of the complete system is this superposed electron state coupled to the two ground states of the cavities $\ket{\text{GS}}_\text{a}$ and $\ket{\text{GS}}_\text{b}$. 
\begin{align}
\ket{\Psi}_{\text{in}}
= \frac{1}{\sqrt{2}}\Big(&\ket{n^{(\text{el})}_\text{a}=1,n^{(\text{el})}_\text{b}=0}\notag\\+&\ket{n^{(\text{el})}_\text{a}=0,n^{(\text{el})}_\text{b}=1}\Big)\ket{\text{GS}}_\text{a}\ket{\text{GS}}_\text{b}\, .
\end{align}

The interaction between an electron in a dot and the corresponding cavity ensures that a coherent state is created, depending on the position of the electron, described by the unitary time evolution operator
\begin{align}
\hat{U}(\tau)=
\hat{\mathcal{D}}_\text{a}
[ \hat{n}^{(\text{el})}_\text{a} \rho(\tau)] \times 
\hat{\mathcal{D}}_\text{b}[ \hat{n}^{(\text{el})}_\text{b}\rho (\tau)]\, ,
\end{align}
where 
$\hat{\mathcal{D}}_{\text{a,b}}[ \xi ]$ are the coherent displacement operators with the associated parameter $\xi$,
and $\rho(t) =-i \lambda t$. (We set $\hbar=1$ here and in the following.)
After some dwell time $\tau$ the state has evolved, and the dot-cavity interaction correlates the two cavities,
\begin{align}
U(\tau)\ket{\Psi}_{\text{in}}= \frac{1}{\sqrt{2}}\Big(&\ket{n^{(\text{el})}_\text{a}=1,n^{(\text{el})}_\text{b}=0}{\ket{\rho(\tau)}_\text{a}\ket{\text{GS}}_\text{b}}\notag\\ +&\ket{n^{(\text{el})}_\text{a}=0,n^{(\text{el})}_\text{b}=1}{\ket{\text{GS}}_\text{a}\ket{\rho(\tau)}}_\text{b}\Big)\text{,}
\end{align}
where $\ket{\rho(\tau)}_\text{a}$ and $\ket{\rho(\tau)}_\text{b}$ are coherent states of the cavities.
The quantum delocalization of the single electron in the double dot leads to a quantum correlation of the two cavities, which indicates the possibility of an entangled state. This state can persist even when the electron leaves the system and the dot state is empty,
\begin{align}
\ket{\Psi}_{\text{out}}= \frac{1}{\sqrt{2}}\ket{0,0}\big({\ket{\rho(\tau)}_\text{a}\ket{\text{GS}}_\text{b}}+{\ket{\text{GS}}_\text{a}\ket{\rho(\tau)}_\text{b}}\big)\text{.}
\end{align}
Notice that in the last step the electron is removed without the knowledge 
of which path it passed through and this operation corresponds to a kind of non-local measurement.

When a second electron subsequently enters the double dot, one can repeat a similar argument starting from the entangled photon state instead of the vacuum. However, for long times, the internal losses of the cavities should be included as well as their energy relaxation and dephasing due to the coupling with the conducting leads via the double dot.
In other words, Fig.~\ref{fig:1}(b) describes just the idealized argument of the entanglement generation process, where it is assumed that the interaction time of the electron and the cavity is short and that decoherence effects of the dots happen much later regarding the timescale of the interaction.

To investigate the possibility of entanglement, we will evaluate two quantities: We prove correlation via the covariance of the Fock occupation numbers and quantum correlation via the violation of the classical Cauchy-Schwarz inequality \cite{Woelk:2002} for the two cavities.
We calculate these quantities by a perturbative expansion in the Keldysh Green's function formalism. To understand the underlying mechanism of the entanglement generation, we compare the results of the double quantum dot system for degenerated levels, namely zero energy difference of the two levels, with those of a large energy level difference. In addition, we discuss the results for a single quantum dot coupled to two cavities with the same dot-cavity coupling constant $\lambda$ but with a dot-lead tunneling coupling of $\sqrt{2}t$, see Fig.~\ref{fig:1}(c).

%
%
%
%

\section{\label{sec:level:3}Theoretical Model}

%
%
\subsection{\label{sec:level3.1}Basic Formalism}

The double quantum dot system with the attached cavities is described by the following Hamiltonian including the electronic and photonic part as well as the electron-photon interaction
\begin{align}
\hat{H} &= \hat{H}_{\text{el}} + \hat{H}_{\text{ph}} + \hat{H}_{\text{int}} \, , \\
\hat{H}_{\text{el}}
&=
\sum_{r=\text{L,R}}
\sum_{k} 
\left( \varepsilon_{kr} -\mu_{r} \right) \hat{c}^{\dagger}_{kr} \hat{c}^{\phantom{\dagger}}_{kr}
+
\sum_{\alpha=\text{a,b}}\varepsilon_{\alpha} \hat{d}^{\dagger}_{\alpha} \hat{d}^{\phantom{\dagger}}_{\alpha} \nonumber \label{eq:Hel}\\
&+ t \sum_{r=\text{L,R}} \sum_{\alpha=\text{a,b}} \sum_k \left( \hat{c}^{\dagger}_{kr} \hat{d}^{\phantom{\dagger}}_{\alpha} + \mbox{h.c.} \right) \, , \\
\hat{H}_{\text{ph}} &= \sum_{\alpha=\text{a,b}}
\omega_{\alpha} \hat{\alpha}^{\dagger} \hat{\alpha}^{\phantom{\dagger}} \, , \\
 \hat{H}_{\text{int}} &= \lambda \sum_{\alpha=\text{a,b}}
 \left( \hat{\alpha}^{\dagger}  + \hat{\alpha}^{\phantom{\dagger}}\right) \hat{d}^{\dagger}_{\alpha} \hat{d}^{\phantom{\dagger}}_{\alpha}
 \, .
\end{align}
In this work we focus on the non-interacting, spinless model for the electronic system \cite{footnote1,Boese:2001}. We disregard the spin degree of freedom, as the interaction with the cavity is spin-independent. The electron spin is therefore not crucial for the entanglement generation process that we focus on here.
$\hat{c}^{\dagger}_{kr}$ and $\hat{d}^{\dagger}_{\alpha}$ with $r=\text{L,R}$ and $\alpha=\text{a,b}$, 
are the creation operators of the electrons in the left and right lead and in the two quantum dots with energy levels $\varepsilon_{\alpha}$.
$\hat{\alpha}^{\dagger}$ and $\hat{\alpha}$ are the creation and annihilation operators of photons with frequency $\omega_{\alpha}$ in cavity $\alpha$. $t \in \mathbb{R}$ is the hopping parameter describing the transport of an electron between dots and leads. 

We focus on the regime of unidirectional transport. 
The voltage that is applied along the system shifts the electrochemical potential of the leads. We consider a symmetric shift in the high voltage limit
\begin{align}
\mu_\text{L}=-\mu_\text{R}=\lim\limits_{eV\rightarrow\infty}eV/2\,.
\end{align}
The Fermi functions of the left and right lead become $f_\text{L}(E)=1$ and $f_\text{R}(E)=0$.
This approximation holds as long as the potential is the largest energy scale 
involved in our model, namely $|eV| \gg \max(k_\text{B} T, \Gamma, \Delta\varepsilon, \omega_0, \eta)$. Here $k_\text{B}T$ is the temperature of the leads, and $\eta$ is the damping parameter, characterizing the cavity losses (see below). The origin of the dot energy levels is chosen to be in the middle of the electrochemical potentials of the two leads, i.e.\ $\bar{\varepsilon}=0$. Dot energies are therefore specified only by the level difference, i.e.\ $\varepsilon_\text{a,b}=\pm\Delta\varepsilon$.

Let us note that we assume identical dot-lead couplings in Eq.~(\ref{eq:Hel}). If dot levels are degenerate, $\Delta\varepsilon=0$, and dot-electrode couplings differ, the transmission function typically shows a dip at energies in close vicinity to $\bar{\varepsilon}$ instead of the peak arising at all identical couplings. We explain in the Supplemental Material \cite{SM}, why a small asymmetry in electrode-dot couplings is not expected to affect our main results. The essential argument is that, since we are assuming the high voltage limit, a local singularity of transport properties at $\bar{\varepsilon}$ is averaged out. The singularity thus does not lead to a discontinuous behavior of the integral of the total electron flux or any other quantities, like Feynman diagrams, derived from integrals over energy \cite{Ryzhik:2014}.

To determine the electronic transport through the system in the absence of electron-photon interaction, 
we calculate the unperturbed electronic Green's functions associated with the electronic part $\hat{H}_{\text{el}}$ of the Hamiltonian, using the diagrammatic Keldysh technique and applying the wide-band approximation for the leads \cite{Shahbazyan:1994,Kubala:2002}. In this way we obtain the broadening $\Gamma$ of the electronic levels. These electronic Green's functions represent our bare propagators in the perturbative approach, where we expand in terms of the electron-photon interaction.
These unperturbed fermionic Green's functions $G_{\alpha\beta}(t_i,t_j) = -i\avt{ \hat{d}^{\phantom{\dagger}}_{\alpha}(t) \hat{d}^{\dagger}_\beta(t') } $, beyond being in the matrix form of the Keldysh formalism \cite{Rammer:2007}, are also $2\times 2$ matrices regarding the two parallel dots
$\alpha,\beta=\text{a,b}$ \cite{Kubala:2002}.
$T_\text{c}$ is the time-ordering operator with respect to the Keldysh contour.
In a similar way, we define the single-particle bosonic Green's function $D_{\alpha}(t,t')=-i\avt{ \hat{\alpha}(t) \hat{\alpha}^{\dagger} (t') }$ for the two cavities 
and the two-particle function $F_{\alpha\beta}(t,t')=-i\avt{ \hat{\alpha}(t) \hat{\beta}(t) \hat{\alpha}^{\dagger}(t') \hat{\beta}^{\dagger}(t')}$. 
We consider the intrinsic photon losses in the two cavities by a finite broadening $\eta$ of the unperturbed bosonic propagators, denoted as $D^{(0)}_{\alpha}(t,t')$. The explicit form of the unperturbed Green's functions is detailed in the supplemental material, chapter A \cite{SM}.

To demonstrate the entanglement of photons in the two cavities, we calculate the covariance 
\begin{align}
  C=\av{ \hat{n}_\text{a}\hat{n}_\text{b}}-\av{\hat{n}_\text{a}}\av{\hat{n}_\text{b}},\label{Eq.Def.C}
\end{align}
which proves correlation if it is finite, i.e.\ $C\ne 0$, and test the classical Cauchy-Schwarz inequality
\begin{align}
	S=\frac{ \av{{\hat{a}}^{\dagger} \hat{a} {\hat{b}}^{\dagger} \hat{b}} } 
	{ \sqrt{\av{\hat{a}^{\dagger}\hat{a}^{\dagger} \hat{a} \hat{a}}} \sqrt{\av{\hat{b}^{\dagger}\hat{b}^{\dagger} \hat{b} \hat{b}}}
	} \le 1
	\text{,}\label{Eq.Def.CS}
\end{align}
which proves quantum correlation if it is violated. The cavity occupations are defined by the bosonic creation and annihilation operators $\hat{n}_\text{a}=\hat{a}^{\dagger}\hat{a}$ and $\hat{n}_\text{b}=\hat{b}^{\dagger}\hat{b}$, and $S$ is the Cauchy-Schwarz parameter.

To evaluate Eqs.~(\ref{Eq.Def.C}) and (\ref{Eq.Def.CS}), we express the expectation values by Keldysh Green's functions and perform a diagrammatic perturbative expansion in the dot-cavity coupling $\lambda$ up to fourth order. The expectation values in these equations are related to the lesser Green's functions in the limit of equal times of the single-particle and two-particle Green's functions \cite{Rammer:2007}.
In this representation, the average photon number of a single cavity $\av{\hat{n}_{\alpha}}$, the covariance $C$ and the Cauchy-Schwarz parameter $S$ read
\begin{align}
\av{\hat{n}_{\alpha}}=&D_{\alpha}^{<}(t,t),\label{Eq.avn}\\[5pt]
	C
		=& iF_{\text{ab}}^{<}(t,t) +D_\text{a}^{<}(t,t)D_\text{b}^{<}(t,t),
										\label{Eq.C}\\[5pt]
	S
		=&	\frac{F^{<}_{\text{ab}}(t,t)} {\sqrt{F^{<}_{\text{aa}}(t,t) F^{<}_{\text{bb}}(t,t)}}.\label{Eq.S} 
\end{align}

The conditions for entanglement are a nonzero covariance, $C\ne 0$, and a violated classical Cauchy-Schwarz inequality, i.e.\ $S> 1$. Both quantities have no finite contributions up to third order, so we have to calculate them consistently up to the fourth order in $\lambda$.

%
%
\subsection{\label{sec:level3.2}Perturbation Expansion}

For the single-particle bosonic Green's functions in Eqs.~(\ref{Eq.avn}) and (\ref{Eq.C}) we perform a perturbation expansion up to second order and for the two-particle Green's function in Eqs.~(\ref{Eq.C}) and (\ref{Eq.S}) up to fourth order with respect to the dot-cavity interaction Hamiltonian in the interaction picture
\begin{align}
H_{\text{int}}(\tau)=& \lambda \left[	(\hat{a}^{\dagger}+\hat{a})\hat{d}^{\dagger}_\text{a} \hat{d}_\text{a} +(\hat{b}^{\dagger}+
\hat{b})\hat{d}^{\dagger}_\text{b} \hat{d}_\text{b}	\right]_{\tau}\, .\label{Eq.Hint}
\end{align}
To calculate the expectation values of the long chains of field operators occurring in the perturbative expansion, 
we use Wick's theorem, which allows to decompose a contour-ordered string of creation and annihilation operators, derived from a quadratic Hamiltonian, 
into a sum over all possible pairwise products\cite{Rammer:2007}. 

Every product corresponds to unperturbed fermionic $G_{\alpha\beta}(t_i,t_j)$ and bosonic Green's functions $D^{(0)}_{\alpha}(t_i,t_j)$, with $t_i,t_j$ lying on the Keldysh contour. An expansion up to fourth order yields contributions with four bosonic and four fermionic Green's functions, integrated over four different time arguments distributed on the Keldysh contour.
Exemplary we consider the integral 
\begin{align}
I(t,t')&=\lambda^4 \oint\limits_\text{c}  \oint\limits_\text{c}  \oint\limits_\text{c} \oint\limits_\text{c} dt_1 dt_2 dt_3 dt_4\notag\\ & D^{(0)}_\text{a}(t,t_1)  D^{(0)}_\text{b}(t_2,t')D^{(0)}_\text{b}(t,t_3) D^{(0)}_\text{a}(t_4,t') \notag	\\ & G_{\text{ab}}(t_4,t_3) G_{\text{ba}}(t_3,t_4)G_{\text{ab}}(t_1,t_2)G_{\text{ba}}(t_2,t_1) \, ,\label{Eq.Int}
\end{align}
which contributes to the covariance and the classical Cauchy-Schwarz parameter. We can represent these integrals as Feynman diagrams and get three different geometries, depicted in Fig.~\ref{fig:3}, while the latter equation corresponds to the third diagram.
%
%
%
%
%
%
%
\begin{figure}[t!]
\includegraphics[scale=0.6]{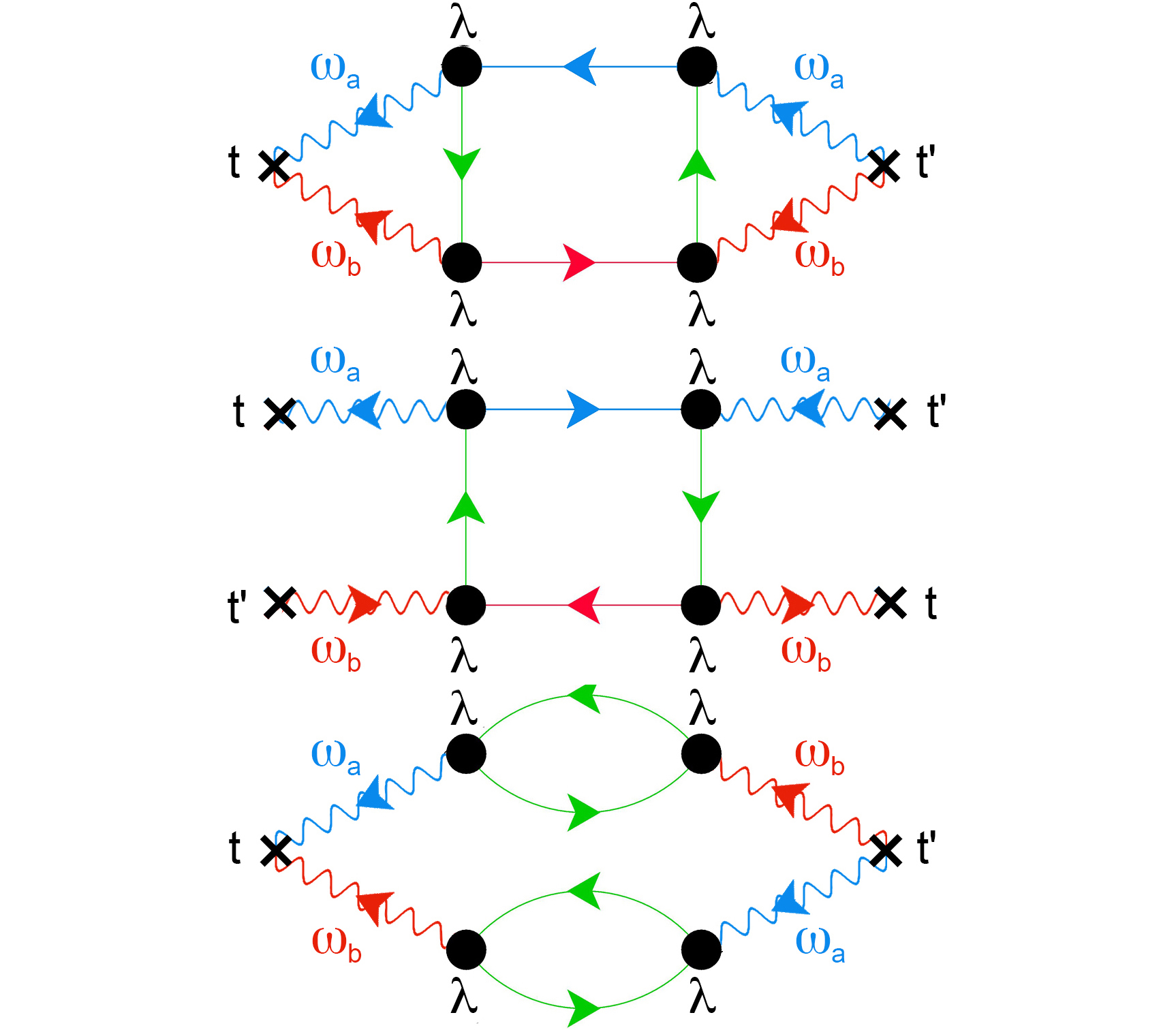}
\caption{Examples of vertex diagrams, which correspond to the integrals of $C$ and $S$. Black dots represent the electron-photon interaction proportional to $\lambda$. Wiggled lines correspond to bosonic Green's functions of the microwave cavities with resonance frequency $\omega_\text{a,b}$, distinguished by the colors red and blue. Straight lines signify fermionic Green's functions of the double quantum dot system. Diagonal ones are indicated by blue and red lines, offdiagonal Green's functions are represented in green.}
\label{fig:3}
\end{figure}
%
%
%
%

Due to the fourth order of the perturbative expansion we get four interaction points proportional to $\lambda$.
This shows that we are dealing with a two-photon process, emitting and absorbing photons in both cavities a and b with corresponding energies $\omega_\text{a,b}$.
The process is described by different types of fermionic interaction in the double quantum dot system.

The complete formulas for $C$ and $S$ are reported in the supplemental material, chapter B \cite{SM}.

%
%
%
%

\section{\label{sec:level4}Results}

To certain limits, we are able to calculate the relevant quantities analytically. 
We consider equal resonance frequencies $\omega_\text{a}=\omega_\text{b}\equiv\omega_0$ and zero temperature of the cavities. Furthermore we focus on the regime of low damping inside the cavities, i.e.\ $\omega_0\gg\Gamma\gg\eta$, and on the high voltage bias limit. Regarding the energy levels of the parallel quantum dots we consider two different cases: First we study the case of two almost equal energy levels and second the case for two strongly differing levels. Finally, we compare these results with the case of a single dot coupled to two cavities at the same time, see Fig.~\ref{fig:1} (c). 

We first determine the average photon number of the single cavities $\av{\hat{n}_{\text{a}}} =\av{\hat{n}_{\text{b}}} =\bar{n}$ up to second order and the corresponding fluctuations $\delta n^2=\av{ \hat{n}^2}-\bar{n}^2$ up to fourth order. The latter quantity can be easily computed from the knowledge of $\bar{n}$ and $\av{\hat{\alpha}^{\dagger}\hat{\alpha}^{\dagger} \hat{\alpha} \hat{\alpha}}$ (viz.\ $F^{<}_{\alpha\alpha}(t,t)$). Then one can analyze the behavior of the Fano factor, defined as $F=\delta n^2 / \bar{n}$. Finally, we calculate the covariance $C$ and the Cauchy-Schwarz parameter $S$ both up to fourth order in $\lambda$. 
Due to the symmetries in the system, the results for both cavities are the same.

\renewcommand{\arraystretch}{2.5} 
\begin{table}
\begin{ruledtabular}
  \begin{tabular}{|c|c|c|c|}  
   & \hspace{-16pt}DQD, $\Delta\varepsilon=0$ & \hspace{-16pt}DQD, $\Delta\varepsilon\gg\Gamma$ & \hspace{-16pt} single dot \\  
   \hline
   $\bar{n}$      	&  $n_\text{0}$	& $2n_\text{0}$ & 2$n_\text{0}$ \\  
   \hline
   $\delta n^2$ 	&$n_\text{0} \left(1- n_\text{0}/4 \right)$& 2${n}_\text{0} \left(1+ {n}_\text{0} \right)$ 		&  $2{n}_\text{0}\left(1+4 {n}_\text{0} \right)$ 													\\ 
   \hline 
   $F$ & $1-n_\text{0}/4$ & $1+{n}_\text{0}$ &  $1+4 {n}_\text{0}$ \\  
    \hline
    $C$ & $n_\text{0}^2 /2$ & 0 & $8{n}^2_\text{0}$  \\
    \hline
   $S$ & 2 & $ 2/3$ & 1  
  \end{tabular}
    \caption{Results for the average photon number of a single cavity $\bar{n}$, 
    fluctuations $\delta n^2$, 
    Fano factor $F$, 
    covariance $C$ 
    and Cauchy-Schwarz parameter $S$,
    for three different cases: the double quantum dot (DQD) with zero or large level spacing and a single quantum dot coupled to two cavities. Here $n_\text{0}= \pi^2 \left(\frac{\lambda}{\omega_0}\right)^2\left(\frac{\Gamma}{\eta} \right)$.
   }
      \label{tab:table01} 
    \end{ruledtabular}
\end{table}
The average occupation for the single dot and the double dot with large level spacing are equal, since we calculate the average occupation only up to second order. The result of the double dot with two almost equal energy levels, i.e.\ $\Delta\varepsilon\ll\Gamma$, acting as single-electron splitter, is half the size of the case $\Delta\varepsilon\gg\Gamma$ and the single-dot system. 
In the case of zero or small level spacing the electronic Hamiltonian, Eq.~(\ref{eq:Hel}), can be written in form of an effective single-dot problem. Compared to the real single-dot case the dot-lead coupling parameter $t$ is renormalized to $t\rightarrow t / \sqrt{2}$. With $\Gamma\propto |t|^2$ this renormalization enters as a factor two in the bosonic occupation of the cavities \cite{Kubala:2002}.
Furthermore we checked that the result for the single dot coupled simultaneously to two cavities coincides with previous results in the limit, in which the induced damping associated with the electron-boson interaction is smaller than the intrinsic damping of the cavities $\eta$ \cite{Mitra.2004,Pascal:2014}. 

According to table~\ref{tab:table01}, for the fluctuations and the corresponding Fano factor of the double-dot system we find a sub-Poissonian behavior in the regime, where we have quantum interference in transport through the double dot, whereas we obtain a super-Poissonian behavior for the other two cases. The sub-Poissonian behavior corresponds to a photon anti-bunching in the local cavity. Let us emphasize that the nature of the interaction already appears at the level of a single-cavity quantity, namely the local fluctuations of the photons in a cavity.
This fact can already be seen at two trivial examples, since the sub-Poissonian behavior occurs both for the entangled bosonic states in the Fock occupation $\ket{\Psi}_{pq} \propto \left| n_\text{a}=p, n_\text{b}=q \right>  +  \left| n_\text{a}=q, n_\text{b}=p \right> $ with $p,q\in \mathbb{N}$ or in the 
coherent state basis $\ket{\Psi}_{z_1z_2} \propto \left| \xi_\text{a}=z_1,  \xi_\text{b}=z_2 \right>  +  \left|  \xi_\text{a}=z_2,  \xi_\text{b}=z_1\right> $, with $\left| \xi \right> $ being a coherent state and $z_1,z_2\in\mathbb{C}$. 

For the covariance $C$ we find a finite, positive value for the double quantum dot with two equal energy levels, which verifies a correlation of the photons in the single cavities.
The covariance for a large level spacing vanishes, meaning that there is no correlation.
This result is expected for the case of two separated electron pathways.
Notice however that a finite covariance also arises in the case of a single dot simultaneously coupled to two cavities, see Fig.~\ref{fig:1}(c). This can be interpreted as classical correlation, as 
we have a single photon emitter coupled to both cavities.

Finite covariance proves correlation but does not indicate quantum correlation (entanglement). 
To distinguish classical and quantum correlations, we calculated the classical Cauchy-Schwarz parameter $S$ of Eq.~(\ref{Eq.S}), which we cast in the following form
\begin{align}
S=& 
\frac{  \left|  \bar{n}^2 + C  \right| }{ \left| \bar{n}\left(  \bar{n}-1 \right)  + \delta n^2 \right| }
=
\frac{ \left|  \bar{n} + \frac{C}{\bar{n}} \right|  }{ \left| \bar{n} + F - 1 \right| }\, .
\label{Eq.S_F_n}
\end{align}
The expression states that a violated Cauchy-Schwarz inequality ($S>1$) occurs, if a finite and positive covariance is combined with a sub-Poissonian ($F<1$) behaviour.
As reported in table~\ref{tab:table01}, the Cauchy-Schwarz inequality for vanishing level spacing is clearly violated.
This confirms the quantum entanglement of the photons in the two distant microwave cavities, if the energy levels of the two dots are sufficiently close to each other, viz.\ the electron is delocalized over the two dots, when it flows from one lead to the other. For strongly differing energy levels of the dots, the classical Cauchy-Schwarz inequality is not longer violated as $C=0$ (uncorrelated systems).
As a sanity check, for the single dot with $C>0$ and $F>1$ (super-Poissonian) we find that the classical Cauchy-Schwarz inequality is not violated but reaches the maximum classical value.

Let us finally discuss the role of decoherence in the system.\\
Intrinsic contributions stem from losses in the cavities. Therefore, we assume high quality cavities with an intrinsic damping that is smaller than the broadening of the electronic levels $\eta \ll \Gamma \ll \omega_0$. We expect the photon production rate to be proportional to the flow of electron through the dots. If the cavities lose energy at a rate which is  
faster than the rate at which photons are created, this will of course destroy entanglement. \\
Another source of decoherence arises from the stochastic nature of electron tunneling. The granular electron flow cannot generate a pure quantum state of the photons in the two cavities, but it will create an entangled mixed state.

We propose our setup as a proof of concept to realize bosonic quantum correlations mediated by single-electron transport. Although several parameters were assumed to be equal, we expect our results to be relevant for carefully chosen experimental settings. Since the covariance $C$ and the Cauchy-Schwarz parameter $S$ are continuous functions of the parameters of the Hamiltonian, we expect our idealized proposal to be robust to small variations and hence to be realizable. A long-term perspective may be the creation of pure entangled states. For this objective a time-dependent control of the electron occupations of the dots may need to be implemented along the lines discussed in \cite{Hofheinz:2008dq,Hofheinz:2009ba}.

%
%
%
%

\section{\label{sec:level5}Summary}

We studied single-electron transport through a parallel double quantum dot, each dot being coupled to a separate microwave cavity. We showed in a simple scheme that for degenerate dot energies, when quantum interference between the two transport pathways is most pronounced, the delocalized electron entangles the photons in two separated microwave cavities. Using the Keldysh Green's function method and a perturbative expansion up to fourth order in the dot-cavity interaction strength, we demonstrated quantum correlations between the cavities through a nonzero covariance and the violation of the classical Cauchy-Schwarz inequality.
Due to the complexity of the calculations we presented analytical results only for the perfectly symmetric case, with both quantum dots exhibiting equal energy levels. But we expect that our findings are still valid, if the  degeneracy is lifted and the system becomes slightly asymmetric.
For too large level detunings or a single dot coupled to two cavities, we have shown that the photons in the different cavities cannot be entangled.

%
%
%
%

 \begin{acknowledgments}
The work was supported by the Deutsche Forschungsgemeinschaft (DFG, German Research Foundation) through Project-ID 32152442 - SFB 767 and Project-ID 25217212 - SFB 1432, and by the German Excellence Strategy via the Zukunftskolleg of the University of Konstanz.
\end{acknowledgments}


\nocite{*}

\bibliography{apssamp}

\end{document}



\title{Supplemental Material for "Quantum-correlated photons generated by nonlocal electron transport"}

%
\author{Felicitas Hellbach}
\affiliation{Fachbereich Physik, Universit{\"a}t Konstanz, 78457 Konstanz, Germany}
\author{Fabian Pauly}
\affiliation{Institute of Physics, University of Augsburg, 86135 Augsburg, Germany}
\author{Wolfgang Belzig}
\affiliation{Fachbereich Physik, Universit{\"a}t Konstanz, 78457 Konstanz, Germany}
\author{Gianluca Rastelli}
\affiliation{Fachbereich Physik, Universit{\"a}t Konstanz, 78457 Konstanz, Germany}
\affiliation{INO-CNR BEC Center and Dipartimento di Fisica, Universit{\`a}  di Trento, 38123 Povo, Italy}

\date{\today}

\maketitle


\section{\label{sec:levelA1} Green's Functions}
\subsection{\label{sec:levelA1.1}Fermionic Subsystem}
We determine the bare fermionic Green's functions of the double quantum dot system and the bare bosonic Green's functions of the single cavities separately, using the diagrammatic Keldysh Green's function technique. The fermionic Keldysh Green's function is a $2 \times 2$ matrix in Keldysh space

\begin{align}
\check{G}(E)=&\begin{pmatrix}
G^{}(E) & G^{<}(E) \\ G^{>}(E) & \tilde{G}(E)
\end{pmatrix}\,,\label{SM.Eq.check-GF}
\end{align}
representing the Fourier transformations of the time-ordered $G(t,t')$, anti-time-ordered $\tilde{G}(t,t')$, lesser $G^<(t,t')$ and greater $G^>(t,t')$ Green's functions. The energy-dependent Keldysh Green's function are determined by the energy-dependent retarded and advanced Green's functions, that are calculated via the Dyson equation and the self-energy. 
Each component in turn is again a $2 \times 2$ matrix in the subspace of the parallel double quantum dot, i.e. for the time-ordered Green's function
\begin{align}
{G}(E)=&\begin{pmatrix}
G_{\text{aa}}(E) & G_{\text{ab}}(E) \\ G_{\text{ba}}(E) & G_{\text{bb}}(E) \\
\end{pmatrix}\,,\label{SM.Eq.GF-time-ordered}
\end{align}
where $\alpha,\,\beta=\text{a,b}$ in $G_{\alpha\beta}(E)$ refer to the upper or lower dot, respectively, see Fig.~1(a) in the main text.

We consider the Keldysh Green's function in the high voltage bias limit. For this reason the Fermi functions of the left and right lead become $f_\text{L}(E)=1$ and $f_\text{R}(E)=0$. $\Gamma\propto|t|^2$ is the broadening of the electronic levels from the coupling to the electrodes, which is treated symmetrically for both sides. Under these conditions the greater and lesser Green's functions read 
\begin{align}
G^{<}(E)=&-G^{>}(E)	\\[5pt]
%
				=&  \frac{i\Gamma }{(E^2-\Delta\varepsilon^2)^2+4\Gamma^2E^2 } 
					\begin{pmatrix}
						 	(E+\Delta\varepsilon)^2  & (E^2-\Delta\varepsilon^2) \\[5pt]
						 (E^2-\Delta\varepsilon^2) &  (E-\Delta\varepsilon)^2
					\end{pmatrix}\,  , \label{SM.Eq.G+-}
\end{align}
and the time-ordered and anti-time-ordered Green's functions are
\begin{align}
G(E)=&-\tilde{G}(E)\\[5pt]
				=&  \frac{1 }{(E^2-\Delta\varepsilon^2)^2+4\Gamma^2E^2 } 
											\begin{pmatrix}
									 (E-\Delta\varepsilon)(E+\Delta\varepsilon)^2+2\Gamma^2E& 
									-2\Gamma^2E \\
									 -2\Gamma^2E  & 
									(E-\Delta\varepsilon)^2(E+\Delta\varepsilon)+2\Gamma^2E
											\end{pmatrix}\,  .
\end{align}

%
%

\subsection{\label{sec_levelA1.2}Bosonic Green's functions}

We define the time-dependent bosonic Keldysh Green's functions as
\begin{align}
\check{D}(t,t')=&
\begin{pmatrix}
D(t,t') &  D^{<}(t,t') \\ D^{>}(t,t') & {\tilde{D}}(t,t')
\end{pmatrix}\\[10pt]
=& -i
\begin{pmatrix}
  \avt{	\alpha^{\dagger}(t')\alpha(t)	} & \av{\alpha^{\dagger}(t')\alpha(t)} \\  
\av{\alpha(t)\alpha^{\dagger}(t')} & \avat{	\alpha^{\dagger}(t')\alpha(t)	} 
\end{pmatrix}\, ,
\end{align}
with ${T_\text{c}}$ and $\tilde{T_\text{c}}$ being the time-ordering and anti-time-ordering operator. For a time independent Hamiltonian, as considered here, $\check{D}(t,t')$ and all of its components will just depend on the time difference $t-t'$. To obtain the unperturbed bosonic Green's functions in energy space, we calculate the Fourier transformation with respect to the time difference, yielding
\begin{align}
{D_{\alpha}^{<}}^{(0)}(E)=& \frac{ -in_{\alpha}\eta}{(E-\omega_{\alpha})+\eta^2} \label{SM.Eq.D+-} \, ,\\
{D_{\alpha}^{>}}^{(0)}(E)=& \frac{-i(n_{\alpha}+1)\eta}{(E-\omega_{\alpha})+\eta^2}\label{SM.Eq.D-+} \, , \\
D^{(0)}_{\alpha}(E)=& \frac{E-\omega_{\alpha}-i\eta(2n_{\alpha}+1)}{(E-\omega_{\alpha})^2+\eta^2}\label{SM.Eq.D++} \, , \\
\tilde{D}^{(0)}_{\alpha}(E)=&\frac{-(E-\omega_{\alpha})-i\eta(2n_{\alpha}+1)}{(E-\omega_{\alpha})^2+\eta^2} \, . \label{SM.Eq.D--}
\end{align}
If the cavity is coupled to an external bath, as for instance a transmission line into which photons from the cavity escape, the bosonic Green's functions contain a finite imaginary part $\pm i\eta$, as used in the expressions above, which describes the damping of the cavity resonator.

We treat the cavities at zero temperature, such that the initial photon number $n_{\alpha}$ is zero. 
Since we are interested only in the lesser functions, compare Eqs.~(\ref{SM.Eq.avn}) to (\ref{SM.Eq.S}) and $n_{\alpha}=0$, in the final formulas of the integrals only the unperturbed time-ordered and anti-time-ordered bosonic Green's functions
\begin{align}
D^{(0)}_{\alpha}(E)=&\frac{1}{E-\omega_{\alpha}+i\eta},\quad \tilde{D}^{(0)}_{\alpha}(E)=-\frac{1}{E-\omega_{\alpha}-i\eta}
\end{align}
 will contribute.

%
%
%
%

\section{\label{sec:levelA2} Perturbation Expansion}
\begin{figure}
    \begin{center}
		\includegraphics[scale=0.44]{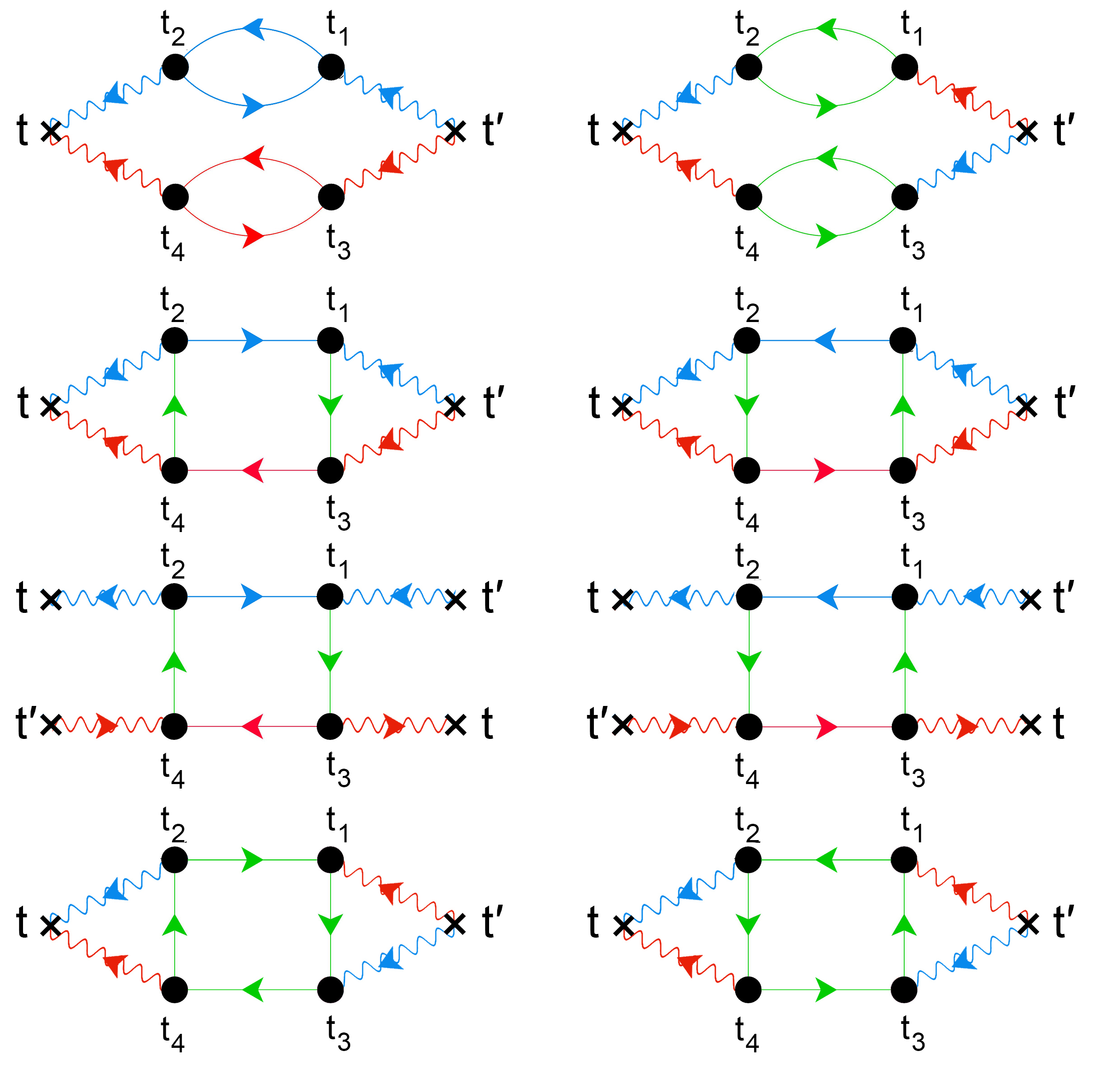}
	\end{center}
	\caption{Eight different vertex diagrams, representing the eight integrals that determine the covariance $C$. The black dots represent the electron-photon interaction, which is proportional to $\lambda$, at time arguments $t_1$ to $t_4$, lying on the upper ($+$) or lower ($-$) branch of the Keldysh contour. The wiggled lines correspond to the bosonic Green's functions of the microwave cavities with resonance frequency $\omega_\text{a,b}$, and the straight lines to the fermionic Green's functions of the double quantum dot system.}
	\label{SM:fig:1}
\end{figure}

For each Green's function in
\begin{align}
\av{\hat{n}_{\alpha}}=&D_{\alpha}^{<}(t,t),\label{SM.Eq.avn}\\[10pt]
	C
		=& \left(iF_{\text{ab}}^{<}(t,t) +D_\text{a}^{<}(t,t)D_\text{b}^{<}(t,t)\right),
										\label{SM.Eq.C}\\[10pt]
	S
		=&	\frac{F^{<}_{\text{ab}}(t,t)} {\sqrt{F^{<}_{\text{aa}}(t,t) F^{<}_{\text{bb}}(t,t)}}.\label{SM.Eq.S} 
\end{align}
we perform a perturbative expansion with respect to the dot-cavity interaction Hamiltonian 
\begin{align}
H_{\text{int}}(\tau)=& \lambda \left[	(\hat{a}^{\dagger}+\hat{a})\hat{d}^{\dagger}_\text{a} \hat{d}_\text{a} +(\hat{b}^{\dagger}+
\hat{b})\hat{d}^{\dagger}_\text{b} \hat{d}_\text{b}	\right]_{\tau}\, .\label{SM.Eq.Hint}
\end{align}
$\hat{d}^{\dagger}_{\alpha}$ and $\hat{d}_{\alpha}$ are the creation and annihilation operators of the electrons in the two quantum dots with energy levels $\varepsilon_{\alpha}$. $\hat{\alpha}^{\dagger}$ and $\hat{\alpha}$ are the creation and annihilation operators of photons with frequency $\omega_{\alpha}$ in cavity $\alpha$. In both cases $\alpha=\text{a,b}$.
Since zeroth and second order contributions to $C$ and $S$ are zero, we calculate up to fourth order. Regarding Eq.~(\ref{SM.Eq.Hint}), we get integrals over a long chain of bosonic and fermionic creation and annihilation operators. We can transform the relevant expectation values into a sum over products of four time-dependent fermionic and four bosonic Green's functions using Wick's theorem. The integrals can be represented as Feynman diagrams. Exemplary, we display the diagrams that determine the covariance in Fig.~\ref{SM:fig:1}. To distinguish the different types of Green's functions occurring, we use straight lines for the fermionic and wiggled lines for the bosonic. Since we have two cavities  and the fermionic Green's functions have four components, describing the parallel double quantum dot subspace, we introduce the following color code:
(i) Light blue color for the bosonic Green's function of the upper cavity ($\check{D}_\text{a}(E)$) and the diagonal element of the fermionic Green's function of the upper quantum dot ($\check{G}_\text{aa}(E)$). (ii) Red color for the bosonic Green's function of the lower cavity ($\check{D}_\text{b}(E)$) and the diagonal element of the fermionic Green's function for the lower quantum dot ($\check{G}_\text{bb}(E)$). (iii) Light green color for the two off-diagonal elements of the fermionic Green's function ($\check{G}_\text{ab}(E)$, $\check{G}_\text{ba}(E)$).

The integrals that contribute to the covariance are visualized in terms of vertex diagrams in Fig.~\ref{SM:fig:1}. 
The structure of the diagrams for the Cauchy-Schwarz parameter $S$ is similar, but they differ in the composition of the components of the Green's functions. Due to the fourth order of the perturbative expansion we get four interaction points proportional to the dot-cavity coupling strength $\lambda$. The diagrams thus describe a two-photon process, involving the emission and absorption of photons in both cavities with energy $\omega_0$, and different types of fermionic interaction in the double quantum dot system. Parametrizing the Keldysh contour through
\begin{align}
\int_\text{c} d\tau =& \int_{-\infty}^{\infty}dt^+ + \int_{\infty}^{-\infty}dt^-\, ,
\end{align}
where $t^+$ are time arguments on the upper branch and $t^-$ on the lower branch of the contour, we transform the integrals from contour time to real time and subsequently sum up the integrand for all possible configurations of the integration variables $t_1$ to $t_4$ on the contour. Regarding Eqs.~(\ref{SM.Eq.avn}) to (\ref{SM.Eq.S}), two time arguments are already fixed due to the choice of the lesser component. This and the fact that the bosonic lesser Green's function of Eq.~(\ref{SM.Eq.D+-}) is zero reduces the number of possible configurations of time arguments on the contour to only one. With the Fourier transformation we express the integrals in energy space.

The average occupation of each cavity is calculated up to second order, where $\bar{n}_\text{a}$ corresponds to the upper (blue) branch of the first diagram in Fig.~\ref{SM:fig:1} and $\bar{n}_\text{b}$ to the lower (red) branch. We consider the limit of equal resonance frequencies $\omega_\text{a}=\omega_\text{b} \equiv \omega_0$. Therefore the expressions for the bosonic Green's functions of Eqs.~(\ref{SM.Eq.D+-}) to (\ref{SM.Eq.D--}) become independent of $\alpha$, i.e.\ equal for both cavities. The average occupations for the two cavities $\alpha=\text{a,b}$ read up to second order
\begin{align}
\bar{n}_{\alpha}= -\lambda^2 \iint d\omega_1\, d\omega_2\,& D^{(0)}(\omega_2-\omega_1)G^<_{\alpha\alpha}(\omega_2)  
G^<_{\alpha\alpha}(\omega_1)\tilde{D}^{(0)}(\omega_2-\omega_1).\label{SM.Eq.avn.alpha}
\end{align}
They become identical for zero level spacing, i.e.\ $\bar{n}_\text{a}= \bar{n}_\text{b}$, since then also the Green's functions of the two dots, the diagonal elements of Eq.~(\ref{SM.Eq.G+-}), are equivalent. For a large level spacing the quantities have to be calculated separately.

To determine the covariance, we need in addition the perturbative expression of $\av{\hat{n}_\text{a}\hat{n}_\text{b}}$ up to fourth order. It is given by
\begin{widetext}
{\small
\begin{align}
& \av{\hat{a}^{\dagger}\hat{a}\hat{b}^{\dagger}\hat{b}}  
=    	   \lambda^4 \hspace{-4pt}	\iiiint \hspace{-2pt}  d\omega_1\, d\omega_2\, d\omega_3\, d\omega_4 
\Big\{ \nonumber \\
&		D^{(0)}(\omega_1-\omega_2) D^{(0)}(\omega_3-\omega_4) \tilde{D}^{(0)}(\omega_1-\omega_2)\tilde{D}^{(0)}(\omega_3-\omega_4) 
		\left[G_{\text{aa}}^{<}(\omega_1)		G_{\text{aa}}^{<}(\omega_2) G_{\text{bb}}^{<}(\omega_3)		G_{\text{bb}}^{<}(\omega_4) +	G_{\text{ab}}^{<}(\omega_3)	G_{\text{ab}}^{<}(\omega_2)			G_{\text{ab}}^{<}(\omega_1)	G_{\text{ab}}^{<}(\omega_4)\right]\notag\\[5pt]
+& 
	D^{(0)}(\omega_1-\omega_2) D^{(0)}(\omega_3-\omega_4) \tilde{D}^{(0)}(\omega_1-\omega_4)\tilde{D}^{(0)}(\omega_3-\omega_2)
		\left[G_{\text{aa}}^{<}(\omega_1)		G_{\text{ab}}^{<}(\omega_2) G_{\text{bb}}^{<}(\omega_3)		G_{\text{ab}}^{<}(\omega_4)  +	G_{\text{ab}}^{<}(\omega_3)	G_{\text{aa}}^{<}(\omega_2)			G_{\text{ab}}^{<}(\omega_1)	G_{\text{bb}}^{<}(\omega_4)\right]\notag\\[5pt]
+&
		D^{(0)}(\omega_1-\omega_2) D^{(0)}(\omega_2-\omega_4) \tilde{D}^{(0)}(\omega_1-\omega_3)\tilde{D}^{(0)}(\omega_3-\omega_4) \notag\\
		&	 G_{\text{ab}}(\omega_2)G_{\text{ab}}(\omega_3) 
			\left[	G_{\text{aa}}^{<}(\omega_1)		G_{\text{bb}}^{<}(\omega_4) +	G_{\text{aa}}^{<}(\omega_4)			G_{\text{bb}}^{<}(\omega_1) + 2\,G_{\text{ab}}^{<}(\omega_1) G_{\text{ab}}^{<}(\omega_4)	\right] \Big\}\, \label{SM.Eq.aabb}.
\end{align}}
\end{widetext}

The two bubble diagrams in the first line of Fig.~\ref{SM:fig:1}, which correspond to the first summand in Eq.~(\ref{SM.Eq.aabb}), are factorizable and
can be separated into their upper and lower branches. This leads to two two-dimensional integrals, which are equal or similar to the integral of the average occupation of Eq.~(\ref{SM.Eq.avn.alpha}). 

\begin{widetext}
To determine the Cauchy-Schwarz parameter, fluctuations and the Fano factor we need to evaluate furthermore 
{\small
\begin{align}
& \av{\hat{\alpha}^{\dagger}\hat{\alpha}^{\dagger}\hat{\alpha}\hat{\alpha}}
=
 2   	\lambda^4 \hspace{-4pt}	\iiiint \hspace{-2pt}  d\omega_1\, d\omega_2\, d\omega_3\, d\omega_4
\Big\{ \nonumber \\
& 2 	 D^{(0)}(\omega_1-\omega_2) D^{(0)}(\omega_2-\omega_3)\tilde{D}^{(0)}(\omega_4-\omega_3)  \tilde{D}^{(0)}(\omega_1-\omega_4) 
			 	G_{\alpha\alpha}(\omega_4)	 G_{\alpha\alpha}^{<}(\omega_1) G_{\alpha\alpha}(\omega_2) G_{\alpha\alpha}^{<}(\omega_3)\notag\\
			+& \left[
			D^{(0)}(\omega_1-\omega_2) D^{(0)}(\omega_3-\omega_4)
			 G_{\alpha\alpha}^{<}(\omega_1)	G_{\alpha\alpha}^{<}(\omega_3)	G_{\alpha\alpha}^{<}(\omega_2)	G_{\alpha\alpha}^{<}(\omega_4)\right] \left[\tilde{D}^{(0)}(\omega_1-\omega_2) \tilde{D}^{(0)}(\omega_3-\omega_4)	+ \tilde{D}^{(0)}(\omega_1-\omega_4) \tilde{D}^{(0)}(\omega_3-\omega_2)\right]\Big\}\, \label{SM.Eq.aaaa},
\end{align}}
\end{widetext}
for $\alpha=\text{a,b}$.

The two-dimensional integral over energy required for the average occupation in Eq.~(\ref{SM.Eq.avn.alpha}) and the separable terms of Eqs.~(\ref{SM.Eq.aabb}) and (\ref{SM.Eq.aaaa}) as well as the four-dimensional integrals of the eight Green's functions in Eqs.~(\ref{SM.Eq.aabb}) and (\ref{SM.Eq.aaaa}) are polynomial in the integration variables and thus in principal solvable. We consider the poles of the functions and perform energy integrations with the help of the residue theorem. Due to the large number and complicated expressions for the poles, algebraic results generally become quite lengthy. To simplify expressions we decided to solve the integrals in the low damping limit, i.e.\ $\omega_0\gg\Gamma\gg\eta$. In the end this yields the results quoted in table~1 of the main text.

%
%
%
%

\section{\label{sec:levelA2} Effect of Arbitrary Tunnel Couplings}

In this section we discuss the effect of arbitrary tunnel couplings on the results for the double dot system without electron-photon interaction,  Fig. (\ref{SM:fig:2}). Specifically, we assume four different tunneling rates 
$\Gamma^\text{L}_\text{a}, \Gamma^\text{R}_\text{a}$ and $\Gamma^\text{L}_\text{b}, \Gamma^\text{R}_\text{b}$.

Let us first study how the saturation current in the high 
bias voltage limit of unidirectional transport depends on the different tunneling rates.
For this purpose, we consider  the transmission function $T(E)$ below. We will show that this function exhibits a singularity when the two dot levels are degenerate, in the sense that $T(E)$ drops to zero instead rising to a maximum, as soon as the system moves away from the high-symmetry point defined as $ \Gamma^L_a\Gamma^R_b = \Gamma^R_a\Gamma^L_b$. However, the saturation current, which is given by the integral of $T(E)$, depends smoothly on this asymmetry due to the energy average.
Hence, a weak asymmetry does not affect this current.

\begin{figure}[h]
    \begin{center}
		\includegraphics[scale=0.44]{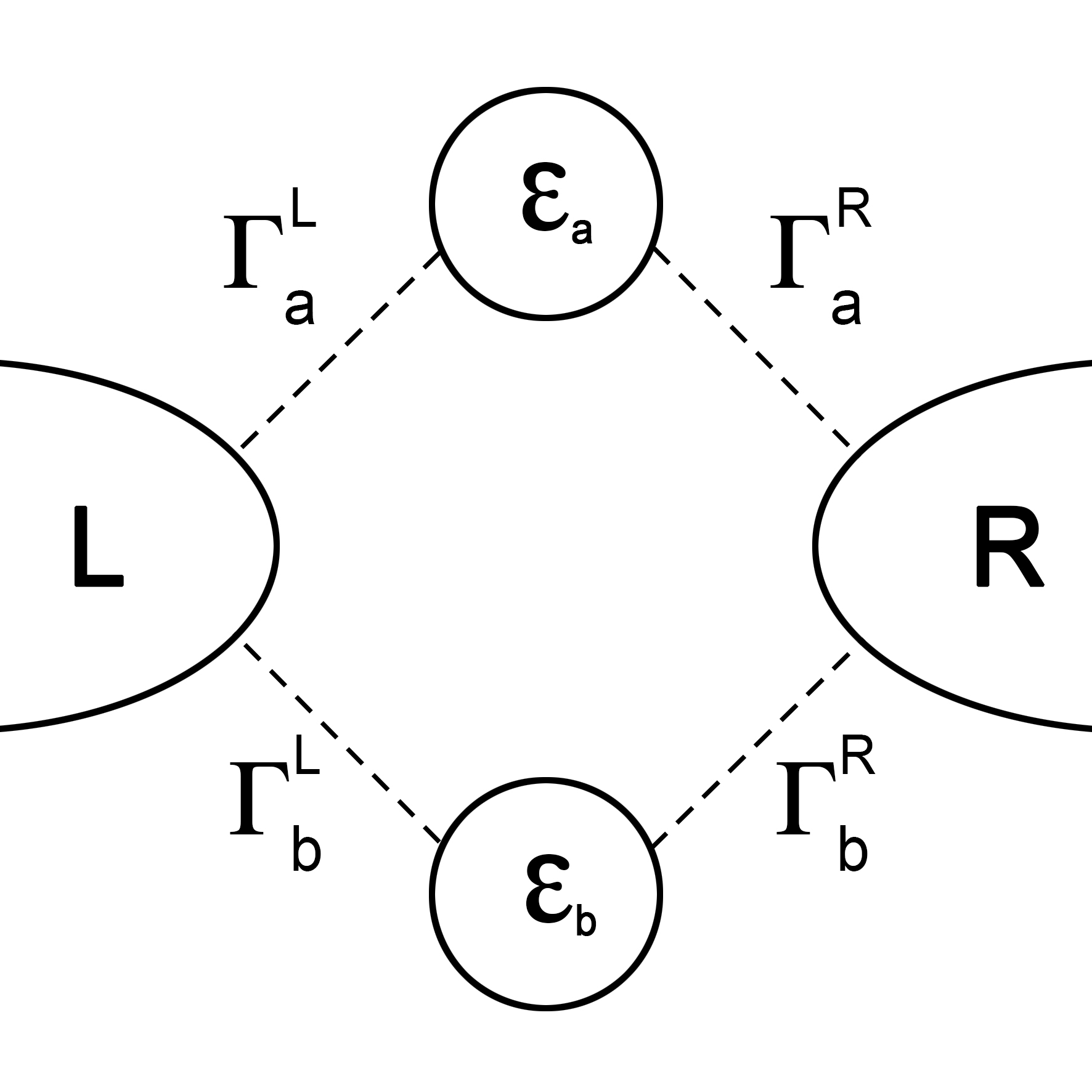}
	\end{center}
	\caption{Parallel double quantum dot with two different energy levels $\varepsilon_\text{a,b}$, coupled to common left and right leads. Tunneling rates between dots and electrodes are described by $\Gamma^\text{L,R}_\text{a,b}$.}
	\label{SM:fig:2}
\end{figure}

The Green's function of the double dot is defined via the $2 \times 2$ matrix
\begin{align}
{G}^\text{r,a}_0(E)
=\left(E \, \mathbb{1}  - {H}_0 -{\Sigma}^\text{r,a}_{\text{L},0}(E)-{\Sigma}_{\text{R},0}^\text{r,a}(E)	\right)^{-1}\, ,
\end{align}
with $(\text{r,a})$ for the retarded and advanced Green's functions, 
$H_0$ the unperturbed Hamiltonian of the isolated levels in the two dots, 
$\Sigma^\text{r,a}_{\text{L},0}(E)$ and $\Sigma_\text{R,0}^\text{r,a}(E)$
the retarded and advanced self-energies due to the left and right leads. The embedding self-energies of the parallel double dot system can be related to the tunneling rates between electrodes and electronic levels $\Gamma^\text{L,R}_\text{a,b}(E)=\pi {\left| t^\text{L,R}_\text{a,b}\right|}^2 \rho_\text{L,R}(E)$, where 
$t^\text{L,R}_\text{a,b}$ are the four different tunneling amplitudes 
and $ \rho_\text{L,R}(E)$ is the density of states of the left and right leads. In the wide-band approximation only the imaginary part of the self-energies is taken into account, and they are assumed to be independent of energy, yielding
\begin{align}
\Sigma_{\text{L},0}^\text{r,a}=& \mp i \Gamma^\text{L} 
= \mp i 
\begin{pmatrix}
\Gamma^\text{L}_\text{a} & \sqrt{\Gamma^\text{L}_\text{a}\Gamma^\text{L}_\text{b}} \\ \sqrt{\Gamma^\text{L}_\text{a}\Gamma^\text{L}_\text{b}} & \Gamma^\text{L}_\text{b}
\end{pmatrix}\, ,\\[10pt]
\Sigma_{\text{R},0}^\text{r,a}=& \mp i \Gamma^\text{R}  
= \mp i 
\begin{pmatrix}
\Gamma^\text{R}_\text{a} & \sqrt{\Gamma^\text{R}_\text{a}\Gamma^\text{R}_\text{b}} \\ \sqrt{\Gamma^\text{R}_\text{a}\Gamma^\text{R}_\text{b}} & \Gamma^\text{R}_\text{b}
\end{pmatrix}\, .
\end{align}
The retarded and advanced Green's functions in matrix form then read
\begin{equation}
G^\text{r,a}_0(E) 
= \frac{1}{\kappa^\text{r,a}({E})} 
\begin{pmatrix}
E-\left({\bar{\varepsilon}}-{\Delta\varepsilon}\right) \pm i\left(\Gamma^\text{L}_\text{b}+\Gamma^\text{R}_\text{b}\right) &\mp i\left(\sqrt{\Gamma^\text{L}_\text{a}\Gamma^\text{L}_\text{b}}+\sqrt{\Gamma^\text{R}_\text{a}\Gamma^\text{R}_\text{b}}\right)\\[10pt] \mp i \left(\sqrt{\Gamma^\text{L}_\text{a}\Gamma^\text{L}_\text{b}}+\sqrt{\Gamma^\text{R}_\text{a}\Gamma^\text{R}_\text{b}}\right) 
& E-\left( {\bar{\varepsilon}}+{\Delta\varepsilon}\right)  \pm i\left(\Gamma^\text{L}_\text{a}+\Gamma^\text{R}_\text{a}\right)
\end{pmatrix}\, ,\label{Eq.Gret-scaled} 
\end{equation}
where the denominator is given by 
\begin{align}
\kappa^\text{r,a}(E)
&=
\left(E- {\bar{\varepsilon}}-{\Delta\varepsilon} \right)
\left(E- {\bar{\varepsilon}}+{\Delta\varepsilon} \right)
-
{\left( \sqrt{\Gamma^\text{L}_\text{a}\Gamma^\text{R}_\text{b}} - \sqrt{\Gamma^\text{R}_\text{a}\Gamma^\text{L}_\text{b}}  \right)}^2
\nonumber \\
&
\pm i\left( \Gamma^\text{L}_\text{a}+\Gamma^\text{R}_\text{a} \right)
\left( E- {\bar{\varepsilon}}+{\Delta\varepsilon} \right)
+
\pm i\left( \Gamma^\text{L}_\text{b}+\Gamma^\text{R}_\text{b} \right)
\left(E- {\bar{\varepsilon}}-{\Delta\varepsilon} \right)
\, 
\end{align}
in terms of the average level position $\bar{\varepsilon}=(\varepsilon_\text{a}+\varepsilon_\text{b})/2$ and the level spacing $\Delta\varepsilon=\varepsilon_\text{a}-\varepsilon_\text{b}$.
The transmission can be derived from the expression
\begin{align}
T(E)=&  4 \text{Tr}\left\{{G}^\text{a}_0(E){\Gamma}^\text{R}{G}^\text{r}_0(E){\Gamma}^\text{L}\right\}\, \label{Eq.Transmission_arbitrary} \, .
\end{align}
Setting $\bar{E}=E-\bar{\varepsilon}$, 
the explicit form of the transmission function becomes
\begin{align}
T(\bar{E})
=&\frac{4{\Gamma}_\text{a}^\text{L}{\Gamma}_\text{a}^\text{R}(\bar{E}+\Delta\varepsilon)^2+4(\bar{E}-\Delta\varepsilon)\left({\Gamma}_\text{b}^\text{L}{\Gamma}_\text{b}^\text{R}(\bar{E}-\Delta\varepsilon)+2\sqrt{{\Gamma}_\text{a}^\text{L}{\Gamma}_\text{b}^\text{L}{\Gamma}_\text{a}^\text{R}{\Gamma}_\text{b}^\text{R}}(\bar{E}+\Delta\varepsilon)\right)}
			{\left[\bar{E}^2-(\Delta\varepsilon)^2-\left(\sqrt{{\Gamma}_\text{a}^\text{L}{\Gamma}_\text{b}^\text{R}}-\sqrt{{\Gamma}_\text{a}^\text{R}{\Gamma}_\text{b}^\text{L}}\right)^2\right]^2+\left[(\bar{E}+\Delta\varepsilon)\left({\Gamma}_\text{a}^\text{L}+{\Gamma}_\text{b}^\text{R}\right)+(\bar{E}-\Delta\varepsilon)\left({\Gamma}_\text{b}^\text{L}+{\Gamma}_\text{b}^\text{R}\right)\right]^2} \, .
\end{align}

Let us now consider the case $\Delta\varepsilon =0$. The transmission can then be expressed as
\begin{align}
T(\bar{E})= \frac{h}{2e}I_\text{sat} \,\, \frac{ 4\bar{\Gamma} \bar{E}^2 / \pi}
			{\left[\bar{E}^2-\delta^2\right]^2+4\bar{\Gamma}^2\bar{E}^2}\, , \label{eq:Te-short}
\end{align}
with
\begin{align}
\delta =& \sqrt{\Gamma^\text{L}_\text{a}\Gamma^\text{R}_\text{b}}-\sqrt{\Gamma^\text{R}_\text{a}\Gamma^\text{L}_\text{b}}\, ,\\
\bar{\Gamma}=& \frac{1}{4} \left( \Gamma^\text{L}_\text{a}+\Gamma^\text{L}_\text{b}+\Gamma^\text{R}_\text{a}+\Gamma^\text{R}_\text{b} \right) \, ,\\
I_\text{sat} =& \frac{2e}{h}\frac{\pi}{\bar{\Gamma}}\left( \sqrt{\Gamma^\text{L}_\text{a}\Gamma^\text{R}_\text{a}}+\sqrt{\Gamma_\text{b}^\text{L}\Gamma_\text{b}^\text{R}}\right)^2\, .\label{eq:Isat}
\end{align}
From Eq.~(\ref{eq:Te-short}) we see that the transmission vanishes at $\bar{E}=0$, i.e.\ $T(0)=0$, if $\delta\ne 0$. 
This dip discontinuously disappears as soon as $\delta=0$. This can be, for instance, the case when all tunnel couplings are equal, as in our study, but the general condition is $\Gamma^\text{L}_\text{a}\Gamma^\text{R}_\text{b}=\Gamma^\text{R}_\text{a}\Gamma^\text{L}_\text{b}$. 

These observations point out a singular dependence of the transmission function on the asymmetry of electronic dot-electrode couplings. However, the critical dependence does not appear in all physical quantities. For example, in the regime of the high voltage bias limit corresponding 
to the unilateral transport $f_\text{L}(E)=1$ and $f_\text{R}(E)=0$, considered in this work, the relevant quantity is the total current which is given by the integral over the transmission.
From Eq.~(\ref{eq:Te-short}) it is possible to show that \cite{Ryzhik:2014:SM}
\begin{align}
I_\text{sat}=
\frac{2e}{h}\int_{-\infty}^{\infty} T(E) \,\, dE
\, , \label{Eq.trans.int}
\end{align}
corresponding to the saturation current. The total current is therefore a smooth function 
of the tunneling rates and depends continuously on the asymmetry of tunneling couplings or tunneling rates, see Eq.~(\ref{eq:Isat}). This result is illustrated in Fig.~(\ref{SM:fig:2}). Since the flux of electrons is largely determined by energies in the interval $[\bar{\varepsilon}- \bar{\Gamma}, \bar{\varepsilon}+\bar{\Gamma}]$, the minimum in the small energy region $[\bar{\varepsilon}-\delta, \bar{\varepsilon}+\delta]$ with $\delta \ll \bar{\Gamma}$ does not matter. 

 \begin{figure}[t!]
    \begin{center}
		\includegraphics[scale=0.25]{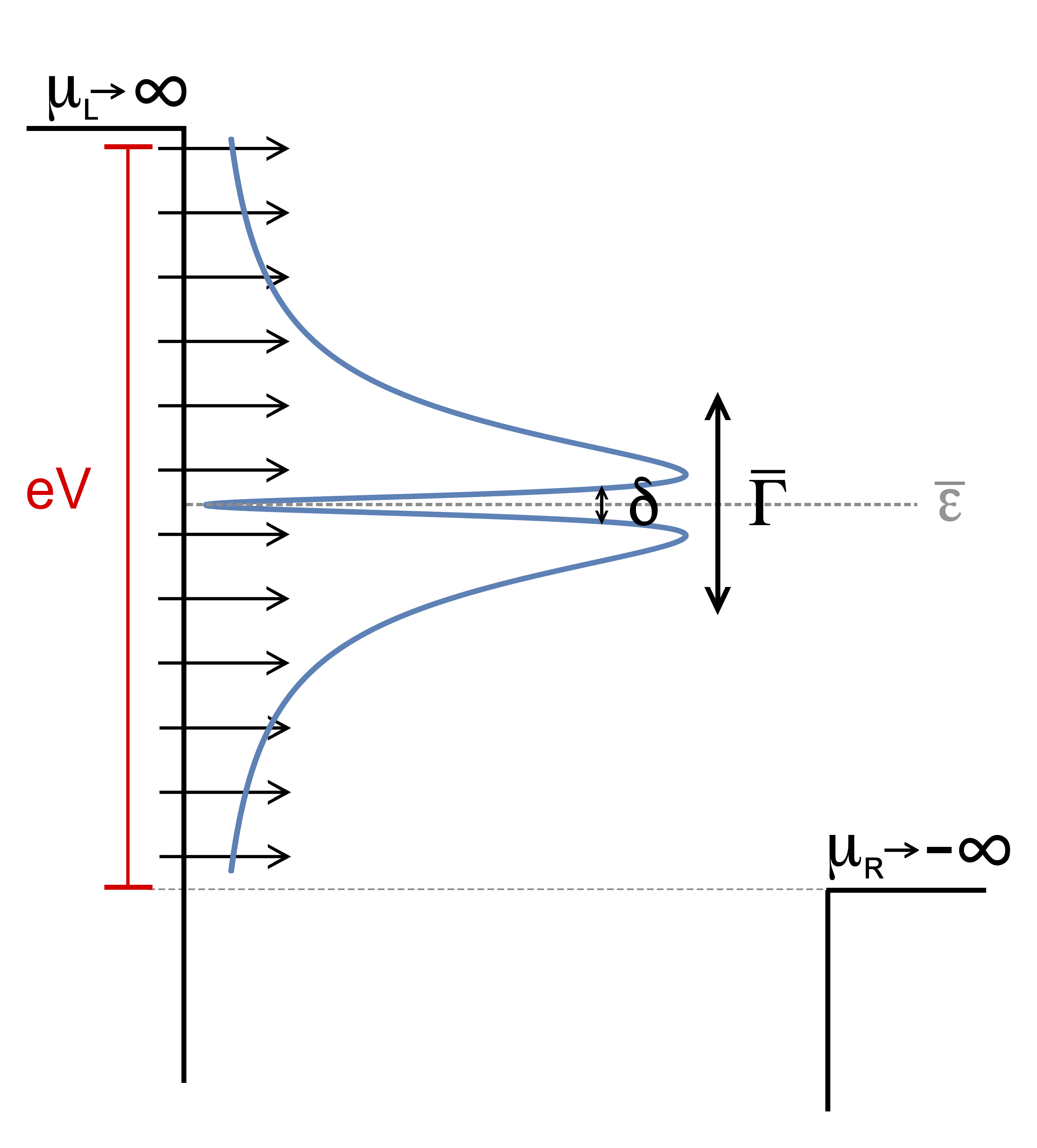}
	\end{center}
	\caption{Elastic transport of electrons (black arrows) through a parallel double quantum dot in the high voltage bias limit, i.e.\  $\mu_\text{L}=-\mu_\text{R}= \lim\limits_{eV\rightarrow\infty}eV/2$. The transmission function is described by the blue curve. The effective flux of electrons is confined by $\bar{\Gamma}$, while the dip associated with the asymmetry of electronic dot-electrode couplings is related to the small energy scale $\delta\ll\bar{\Gamma}$.}\label{SM:fig:2}
\end{figure}

We now discuss the effect of the asymmetry of dot-electrode couplings on the Greens's functions, which are used to construct the diagrams in the perturbation theory to treat the parallel double dot system with electron-photon interaction. For the degenerate case $\Delta\varepsilon=0$, the denominator of the Green's function $G_0^\text{r,a}(E)$ reduces to
\begin{equation}
\kappa^\text{r,a}(\bar{E})
=
\bar{E}^2
-
\delta^2
\pm 4 i \bar{\Gamma}   \bar{E} \, ,
\end{equation}
pointing out that a finite $\delta$ changes the position of the poles.
However, the Green's functions need to be inserted into the integrals defining the diagrams, and the small energy region associated with $\delta$ will generally give a small contribution to the full integral. Therefore we expect a continuous behavior of the diagrams as a function of the asymmetry. In other words, the electrons flowing in the small energy range associated with the region $[\bar{E}-\delta,\bar{E}+\delta]$ with $\delta \ll \bar{\Gamma}$ are expected to yield only a small contribution to the entanglement, when we compare them to the whole flux of electrons. The small energy scale $\delta$ is also irrelevant when we consider inelastic tunneling events, as the resonance frequency of the microwave cavities is much larger, i.e.\  $\omega_0 \gg \bar{\Gamma}$. Overall, these considerations justify our approximation of setting all the tunneling rates equal in order to largely simplify the complex perturbative calculations of the Feynman diagrams.

\bibliography{apssamp-SM}